\documentstyle[epsf]{mn}

\newcommand{\ebv}{ {\it E(B--V)}}

\newcommand{\bb}{\bibitem[]{bla}}
\newcommand{\zm}{ \relax \ifmmode {\rm M_{\odot}} \else {M$_{\odot}$} \fi}

\newcommand{\ang}{$\rm \AA$}

\newcommand{\ea}{{et al.}}

\newcommand{\ha}{H$\alpha$}

\def\lesssim{\mathrel{\hbox{\rlap{\hbox{\lower4pt\hbox{$\sim$}}}\hbox{$<$}}}}
\def\gtrsim{\mathrel{\hbox{\rlap{\hbox{\lower4pt\hbox{$\sim$}}}\hbox{$>$}}}}

\def\ion#1#2{#1$\;${\small\rm\@Roman{#2}}\relax}
 
\newcommand{\lam}{$\lambda$}

\newcommand{\bd}{BD+40$^{\rm o}$4124}

\newcommand{\Av}{$A_{V}$}     
\newcommand{\Rv}{$R_{V}$}

\newcommand{\hab}{Herbig~Ae/Be$\:$}

\newcommand{\etal}{{et al.}}

\begin{document}  

\title[Diffuse bands in YSOs]
{ 
Diffuse interstellar bands in the spectra of massive YSOs
}

\author[Ren\'e D. Oudmaijer ]
{ Ren\'e D. Oudmaijer, Graeme Busfield, Janet E. Drew \\
Imperial College of Science, Technology and Medicine,
Blackett Laboratory, Prince Consort Road,\\ London,  SW7 2BZ, U.K.   \\
}
 \date{received,  accepted}
 
\maketitle
\begin{abstract}
We have compared the $B-V$ colour excess, \ebv , obtained for
a sample of five optically visible massive YSOs both from diffuse interstellar
bands (DIBs) in their spectra and from their optical continuum slopes.
Our targets are HD~200775, \bd , MWC~1080, MWC~297 and MWC~349A.
First,  \ebv\ towards each of the targets is derived by dereddening the 
observed continua to match those of  B-type standard stars.
A survey of DIBs in the spectra of the massive YSOs, and a control field
star, then reveals that the DIBs are significantly weaker in the former
than would be expected based on the total \ebv\ values. This result is 
strengthened by the finding that the DIBs in the control field star, 
HD~154445, have on average the strength expected from its continuum \ebv.  

A rough estimate of the foreground reddening of intervening diffuse
interstellar medium shows it to be smaller than the DIB-\ebv,
implying that at least part of the DIB carriers are formed within the
parental molecular clouds in which the YSOs are embedded.  The
formation efficiency of the DIBs varies strongly however from cloud to
cloud. The DIB-\ebv\ compares favourably with the total \ebv\ towards
\bd, but is almost negligible in the line of sight towards MWC 297.
Despite this general, but not unexpected, deficit we provide evidence
that the DIB at 5849\AA\ is  a good tracer of total extinction in these
lines of sight.

\end{abstract}
 


\section{Introduction}

Still unidentified, diffuse interstellar bands (DIBs) are believed to
be the result of the absorption of starlight by complex molecules or
small dust particles in the interstellar medium.  The weak absorption
features have been well documented by early spectroscopists although
it was not until the 1930s that the interstellar nature of the lines
was recognised by Merrill (1934) and coworkers.  Continuing
investigations showed that the strength of the interstellar absorption
features increased with the star's distance and degree of interstellar
reddening (Merrill, 1936).  Hence the measurement of the strength of
the DIBs is related to path length through the diffuse interstellar
medium (ISM). More recently it has become clear that the strength of
the DIBs correlate with the column density of H{\sc i}, but not with
the H$_2$ column (Herbig 1993, 1995).

Although the line strengths of individual DIBs correlate relatively
well with the foreground reddening in cases where the diffuse ISM is
the main contributor to the extinction, there are particular classes
of sightline where this correlation breaks down.  For example, for
lines of sight through circumstellar material the DIB-derived colour
excesses (\ebv) turn out to be smaller than the total \ebv\ determined
by other means (e.g. Snow \& Wallerstein, 1972; Le Bertre \& Lequeux,
1993 on mass-losing objects; Porceddu, Benvenuti \& Krelowski 1992, on
Be stars).  A similar effect has been found for sightlines into
star-forming regions, with the further complication that not all DIBs
suffer the same reduction (e.g. Meyer \& Ulrich 1984, who studied T
Tau objects; Adamson \ea\ 1991, Taurus dark clouds; Jenniskens,
Ehrenfreund \& Foing 1994, Orion).  These findings suggest that the
physical conditions in these media are such that the carriers
responsible for the DIB absorption do not exist in the same relative
quantities as in the diffuse ISM, which is probably related to the
formation, excitation and destruction of different DIB carriers.  The
concept of different DIB `families' (Krelowski \& Walker, 1987) can be
traced back to this concept.  Some band strengths are well correlated
with each other, but not with the reddening, while other bands
correlate strongly with reddening.

In the present study we describe the behaviour of the DIBs in the lines
of sight towards massive young stellar objects.  Our targets are the
optically visible young stellar object (YSO) MWC 349A and four Herbig Be
stars, objects which are optically bright enough to obtain high
signal-to-noise spectra at intermediate dispersion in the optical.  The
\hab\ stars are widely believed to be associated with circumstellar
envelopes of dust.  
The material in the line of sight towards these young
massive objects is thus a combination of `normal' intervening diffuse
interstellar material, possibly cold material in the clouds in which the
objects have formed, and certainly warm circumstellar material.  It is
hard to estimate the relative contributions to the total extinction of
these media, but observations of the DIBs can in the long run help to
elucidate this problem. 

Here, we investigate whether the DIBs in the spectra of our targets
would imply the same extinction as that derived by other means.
Any difference between the derived
\ebv\ values would set an upper limit to the contribution of
diffuse material  to the line-of-sight extinction.  In the following we will 
derive  \ebv\ by comparing the observed slopes in the target spectra with 
standard star spectra - a `continuum method', and derive \ebv\ from the 
DIB strengths alone using published relations between the EW of DIBs and 
\ebv.

The lay-out of this paper is as follows: In Sec.\ref{OBS} we describe
the observations and data reduction. In Sec.~\ref{EBV} the
determination of the continuum reddening towards the targets is
presented and compared with values in the literature. In
Sec.~\ref{DIBS} the measurement and analysis of the DIBs in the
spectra are presented.  We then compare the DIB derived \ebv\ with
statistical estimates of the colour excess attributable to the
foreground ISM described in Sec.~\ref{foreg}. The paper ends with a
summary.

\section{Observations and data reduction}      
\label{OBS}      
       
In the course of our study of the optical spectra of massive YSOs
(e.g.  Drew \ea\ 1997) we obtained observations on the nights of the
20th to the 22nd June 1994 with the William Herschel Telescope (WHT)
at the La Palma Observatory. The weather throughout the run was good,
yielding a seeing of 0.7 to 1 arcsec.  The twin-beam intermediate
dispersion spectrograph (ISIS) was used with the R1200B and R1200R
gratings on the blue and red arms of the instrument. The blue
detector was a 1124 $\times$ 1124 pixel TEK1 chip, the red detector
was a 1180 $\times$ 1280 EEV6 chip.  The wavelength ranges covered
were 3860--5190\AA\ in the blue arm with a small gap of 120\AA\
between 4270--4390\AA\ . The red arm covered the wavelength range
5840--8390\AA . The dichroic crossed over at 5400\AA .  The
instrumental set-up does not allow for the observation of the
well-known DIBs at 5780 and 5797 \ang. However, the large wavelength
coverage and high sensitivity of the observations make up for this
loss.  For wavelength calibration a copper-neon-argon lamp was
observed at frequent intervals.  The projected entrance slit width was
1 arcsec, which resulted in spectral resolution elements of 0.8\AA \
($\lambda$/$\Delta\lambda$ $\sim$ 8200 at \ha), as determined from arc
line profile fits.  To minimise cosmic ray events in the spectra,
exposure times were limited to 1800s.  The observations
included a spread of spectral type standards which are listed in
Table~\ref{tab:spectypes}. The typical integration times for these
bright stars were of order minutes per setting, resulting in SNR
ratios in the continuum of more than 100.
Table~\ref{tab:objects} lists the YSOs observed.  At each wavelength
setting the spectral coverage obtained was generally a few Angstroms
in excess of 400\AA .  The SNR achieved for each of the objects is
given in the final column in the tables.

\begin{table}    
\caption{The  spectral type standards and atmospheric standards.    
\label{tab:spectypes}    }
\begin{tabular}{|l|l|c|c|}    
\hline    
\hline    
     
Object   & SpTp \\
\hline    
HD 154445& B1.5V  \\
HD 186618& B0.7IV \\
HD 214432& B2.5V((n))  \\
HR 6092  & B5IV	 \\ 
HR 8335  & B3III  \\
\hline    
\hline   
\end{tabular}    
\, \, \\
The spectral types are taken from Walborn (1971)
\end{table}

\begin{table}    
\caption{ Observations of the optically bright YSOs.
The effective wavelength coverage was approximately 400\AA\ per setting    
centred on the wavelengths given in the second column. The co-added SNR    
is given if more than one observation is available.    The {\it B} and {\it V} 
magnitudes are taken from Hillenbrand \ea\ (1992), 
the {\it V} magnitude for MWC 349A  is taken from Cohen \ea\ (1985).
\label{tab:objects}    }
\begin{tabular}{|l|r|c|r|rc}    
\hline    
\hline    
     
Object & {\it B},  {\it V}  &  Wavelength     & Exposure & Continuum \\    
Spectral type&         &  (centre, \AA ) & Time (s) &    SNR   \\    
\hline    
         &                 &           &      \\    
HD 200775  & 7.75,  7.42 &          4600           & 50        &  230     \\    
B2.5e$^1$ &         &  5000           & 50        &  250     \\    
&         &  6055           & 100       &  340     \\    
&         &  6475           & 30        &  140     \\    
&         &  6895           & 100       &  390     \\    
&         &  7315           & 70        &  300     \\    
&         &  7735           & 40        &  220    \\    
        &                 &           &      \\    
\bd & 11.28, 10.54     &  4600           & 1200       &  210     \\    
B3e$^1$ &         &  5000           & 600       &  130     \\    
&         &  6055           & 600       &  190     \\    
&         &  6475           & 630       &  150     \\    
&         &  6895           & 200       &  120     \\    
&         &  7315           & 200       &  135     \\    
&         &                 &           &      \\    
MWC 1080  & 13.08,  11.68 & 4600           & 2648      &   110    \\    
B0$^2$ &         &  5000           & 1000      &   100    \\    
&         &  6055           & 1800      &   200    \\    
&         &  6475           & 200       &   40    \\    
&         &  6895           & 300       &   110    \\    
&         &  7315           & 300       &   135    \\    
         &                 &           &      \\    
MWC 297 & 14.54,  12.17            &  4600           & 9000      &  150     \\    
B1.5Ve$^3$ &         &  5000           & 1800      &  80    \\    
&         &  6055           & 5400      &  340   \\    
&         &  6475           & 2530       &   300    \\    
&         &  6895           & 2100       &  320     \\    
&         &  7315           & 1200      &     305  \\    
&         &  7735           & 500       &    300   \\    
         &                 &           &       \\    
MWC 349A & -, 13.5  
         &  4600           & 5400       &  70     \\    
&         &  6055           & 800        &  120     \\    
&         &  6475           & 270        &  115     \\    
&         &  6895           & 1406       &  205     \\    
&         &  7315           & 2500        & 355    \\    
         &                 &                    &      \\

\hline    
\hline    
\end{tabular}    
\, \\
References to the spectroscopic determinations of the spectral types: \\
$^1$ : Finkenzeller (1985),
$^2$: Cohen \& Kuhi (1979), 
$^3$: Drew \ea\ 1997.
\end{table}

Data reduction was performed using {\sc iraf}. Additional manipulation
made use of both the {\sc iraf} and {\sc starlink-dipso} software
packages. The data were reduced from two-dimensional frames into a
one-dimensional form, through the steps of de-biasing, flat-fielding,
sky subtraction and wavelength calibration.  A first order cubic
spline was used for the wavelength calibration. The fit attained an
rms error in the residuals of less than 0.03\AA.  The extracted
spectra were airmass corrected in order to remove the effects of
atmospheric extinction in the spectra. We applied the appropriately
scaled sum of the wavelength-dependent mean extinction curve for La
Palma (King 1985) and the ``grey'' aerosol component whose magnitude
is measured nightly and archived by the Carlsberg Meridian Circle.
Where necessary, telluric features were removed by dividing the target
star spectra by the corresponding spectra of a suitable comparison
star (either HR 6092 or HR 8335).  The spectra were scaled such that
the depths of the telluric absorption features in the standards match
those in the target spectra.  The main regions affected by telluric
features are 5860--6000\AA, 6270--6330\AA\, 6450--6580\AA\ and
6860\AA\--6950\AA. All wavelengths longwards of 6950\AA\ are
susceptible to sporadic atmospheric features.

\begin{table}
\caption{ Determination of the total \ebv.
The central wavelength of the spectral range is given in column 2.   
Column 3 records the values of the \ebv\ as determined by the H83 law 
derived using each of the standard stars.
Below the lines of measured \ebv s the   
weighted mean of all the \ebv s is presented.   
Previous \ebv\ or \Av\ obtained from the literature are given in the    
final column. \label{tab:ebv}    
}
\footnotesize    
\begin{center}    
\begin{tabular}{|cccc|l|}    
\hline    
\hline    
Object    & $\lambda$ &  \ebv             &Literature& Method \\    
          &           &  B0.7,B1.5,B2.5   &          &              \\    
\hline    
          &           &                   &             &              \\    
HD 200775 & 4600\AA\  & 0.60, 0.45, 0.55  & 0.65 $^1$   & colours \\    
          & 5000\AA\  & 0.75, 0.45, 0.70  & 0.56 $^2$     & colours \\    
          &           &                   & 0.64 $^3$    & SED model. \\     
          &           &  mean: 0.6& 0.57 $^4$    & colours \\     
\hline    
\bd       & 4600\AA\  & 1.00, 0.85, 0.95  & 0.96 $^1$   & colours       \\    
          & 5000\AA\  & 1.15, 0.90, 1.00  & 0.94 $^2$     & colours\\	    
          &           &  mean: 1.0&               \\    
\hline    
MWC 1080  &  4600\AA\ & 1.60, 1.45, 1.45  & 1.71 $^1$   & colours             \\    
          &  5000\AA\ & 1.80, 1.50, 1.75  &\Av=5.42 $^5$  & as here           \\    
          &           &  mean: 1.6&   \\    
\hline    
MWC 297   & 4600\AA\  & 2.95, 2.85, 2.95  & 2.68 $^1$   & colours        \\    
          & 5000\AA\  & 3.05, 2.70, 2.85  & \Av=7.9 $^6$ & NIR slope  \\    
          &           &          mean: 2.9&              \\    
\hline    
MWC 349A  &  4600\AA\ & 2.95, 2.95, 3.00  & \Av=9.9 $^7$ & as here       \\    
          &           &          mean: 3.0& \Av=8.8 $^6$ & NIR  slope  \\    
          &           &                   &          &              \\    
\hline    
\hline    
\end{tabular}    
\, \\
References:
$^1$ - Hillenbrand \etal\ 1992, 
$^2$ - Finkenzeller \& Mundt 1984, 
$^3$ - Voshchinnikov, Molster \& Th\'e 1996
$^4$ - Pfau \etal\ 1987, 
$^5$ - Cohen \& Kuhi 1979, 
$^6$ - McGregor, Persson \& Cohen 1984. 
$^7$ - Cohen \etal\ 1985

\end{center}

\normalsize    
\end{table}

\section{Continuum Determination of Colour Excess.}    
\label{EBV}

Before assessing the relation between the DIB strengths and reddening
towards the target stars, we need to have an estimate of the {\it
total} \ebv\ for each of them. In this section we will derive \ebv\ by
comparing the observed continuum slopes of the spectra with those of
spectral standard stars.  Their spectra were observed during the same
run with the same instrumental set-up. All observations were obtained
with the slit at a parallactic angle to minimise the influence of the
dispersion of the Earth atmosphere. As stated before, the spectra are
corrected for airmass. The procedure is thus independent of the
observed flux from the objects, and only takes into account changes in
the spectral slopes.  It is the same as that used by Cohen \ea\ (1985)
and Drew \ea\ (1997) to derive the reddening to MWC 349A and MWC 297
respectively, both stars are also included in this sample.

The extracted spectra of the target stars are dereddened to find the
best match to the continuum slopes of the standard stars, with \ebv\
as the only free parameter. 
This method implicitly assumes that the intrinsic continuum spectral
energy distributions of the Herbig Be stars in the wavelength ranges
of interest resemble those of the spectral standards closely.  This is
indeed the case for our targets. Spectroscopic determinations of their
spectral types (Table~\ref{tab:objects}) cover the same spectral type
range as our comparison stars (see Table~\ref{tab:spectypes}).  We are
not aware of a spectroscopic spectral type for the extreme emission
line star MWC 349A - but considering its emission line character and
continuum energy distribution, it is widely held to be an early B type star
(see e.g. Cohen \ea\ 1985).  The knowledge of the precise spectral
type is not critical however, since tests with unreddened Kurucz
(1991) models indicate that the difference in slopes between the
extreme spectral types of O9 and B3 corresponds to an error in the
derived \ebv\ of between 0.05 and 0.1, which is much less than our
adopted error in the final \ebv\ values.

Additionally, account needs to be taken of the fact
that the standard stars are lightly reddened themselves.  We derived
their reddening using the observed colours (obtained from the {\sc
simbad} database) and intrinsic colours listed in the tables of
Schmidt-Kaler (1982) for their respective spectral types.  HD~154445
has an \ebv =0.41, while HD~186618 and HD~214432 display a small
amount of reddening of 0.07 and 0.11 respectively.  Dereddening
HD~154445 with respect to HD~186618 and HD~214432, using the same
technique as for the young stellar objects, yielded an \ebv\ difference
consistent with the photometric estimates.

Since we are only interested in the continuum slope, strong
emission lines (in the Herbig Be stars) and absorption lines (in the
standard stars) were snipped out of the spectra to prevent these from
affecting the fitting procedure. 

Once the spectra had been  prepared, the fit procedure was relatively
straightforward: each spectrum was de-reddened for various values of
\ebv\ until an rms minimum between the de-reddened spectrum and the
standard star was reached.  The procedure encountered too few points
upon which to operate satisfactorily in the case of extreme emission line 
stars MWC~1080 and MWC349A.  In these cases, the colour excesses were 
better determined by eye.  

For the determination of the reddening we have used reddening laws
from two sources, the mean \Rv\ = 3.1 galactic extinction law compiled
by Howarth (1983, H83), and also the \Rv\ dependent analytic
prescriptions due to Cardelli, Clayton \& Mathis (1989, CCM).  The red
settings proved not sensitive enough to return useful extinction
estimates, and only the results for the 4600 and 5000 \ang\ settings
are listed in Table~\ref{tab:ebv}.

\subsection{Results}

 The YSO colour excesses, derived using the H83 law, are set out in
 Table~\ref{tab:ebv}. These include the correction for reddening of
 the standard stars.  Below the lines of measured \ebv s, the weighted
 mean of all the \ebv s for a given star is presented. As we associate
 3$\sigma$ errors of 0.15 and 0.2 with the 4600 and 5000 \ang\
 settings respectively, the weights applied to them in obtaining the
 mean were chosen to be in the ratio 4:3.  It is worthy of note that
 the derived \ebv\ of MWC~349A and MWC 1080 are in good agreement with
 the literature values despite the extreme emission line character of
 their spectra.

\label{EBV:RES}

In principle the CCM law
follows - as the H83 law - the van de Hulst no. 15 law (van de Hulst, 1949) 
closely for \Rv = 3.1. The main advantage of the CCM law over
the H83 law is that the dependence on the ratio of total to selective
reddening may be investigated. It turned out that the \Rv\ sensitivity
in our spectra is very small; fits using different values for \Rv\
yielded similar fit qualities and results as those with \Rv\ = 3.1. On
the whole, the CCM law results were within 1$\sigma$ of the results
outlined in the table. Exceptions to this were obtained for the highly
reddened objects MWC 349A and MWC 297 where the CCM law yielded lower
\ebv\ values by as much as 0.45 in the 4600 \AA\ setting.  This is
because the CCM law is somewhat steeper in the 4600 \AA\
wavelength range than the H83 law, a difference that is amplified for
highly reddened objects.

\section{DIB determination of colour excess}
\label{DIBS}

\begin{figure}
\mbox{\epsfxsize=0.45\textwidth\epsfbox[56 175 326 518]{
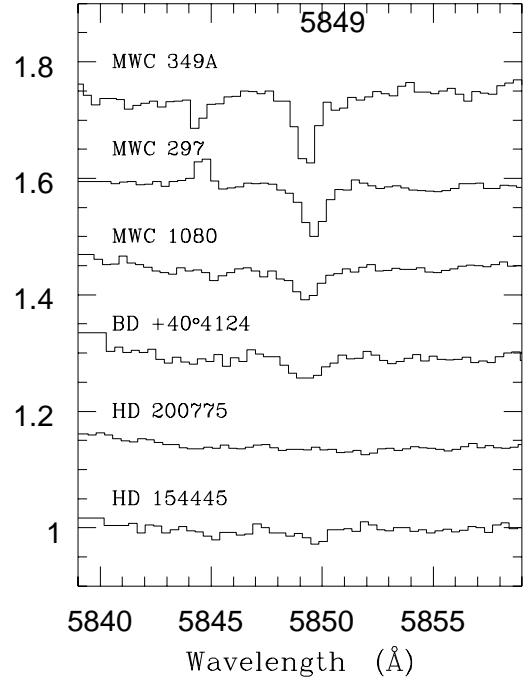}}
\mbox{\epsfxsize=0.45\textwidth\epsfbox[56 175 326 518]{
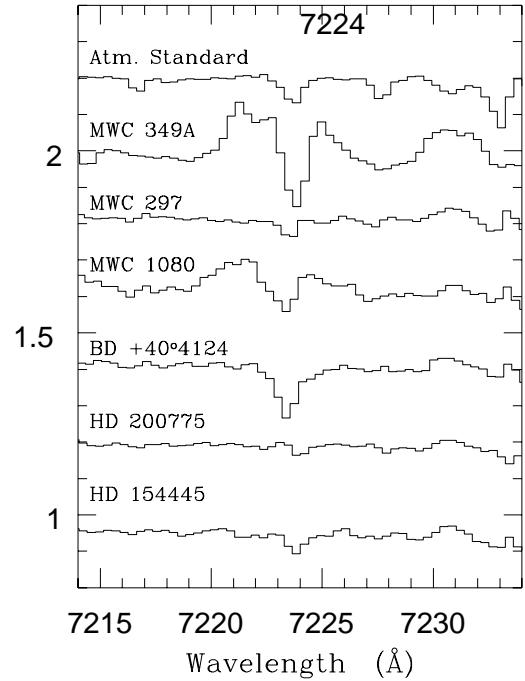}}
\caption{
Continuum normalized spectra around the $\lambda$5849 and
$\lambda$7224 DIBs. The stars are plotted in order of decreasing \ebv\
from top to bottom.  The reddened standard star HD 154445 is shown as
comparison in both figures.  For the $\lambda$7224 DIB, the
atmospheric standard HR 8335 is shown to illustrate results of the
correction of telluric absorption lines.
\label{fig:dibs}
}
\end{figure}

We selected all DIBs listed by Herbig (1995) which appear in the
wavelength range covered by our data.  The DIBs in Herbig's original
list with a normalised equivalent width $<$ 8 m\AA\ are below our
detection capabilities and are not considered.  To this list we added
stronger lines listed by Jenniskens \& D\'esert (1994, hereafter JD)
but not by Herbig (1995).  The final list  comprises 80 DIBs of
which 13 are from the JD catalogue.

With the search list  in hand, the spectra were examined independently
for DIBs by two of us to minimize the inevitable
subjective element.  In general, the two independent assessments yielded
identifications and equivalent widths that were in agreement at the
10-15\% level.  In those cases where extreme differences appeared,
such as non-identifications or very different equivalent widths,
the spectra were inspected again.  Often, the combination of faint
features and the presence of emission lines in the spectral
neighbourhood caused the differences. The measured FWHM are in
accordance with their listed values.

Because of the emission line nature of the objects, many lines were
hampered by blends with emission lines. The spectral resolution of the
observations is in most cases enough to sample the emission lines
sufficiently to enable us to distinguish them from the DIBs.  After
carefully selecting the least affected lines, a sample of 18 DIBs
remained. These lines are presented in Table~\ref{tab:dibs}.  The
reddened spectral type standard HD~154445 displays a number of DIBs
and is also included in the investigation as a fortuitous control.
Examples of two DIBs are given in Fig.~1.
  
In the following, we will discuss the reddening as traced by the DIBs,
and exploit the concept of a DIB-\ebv .  These are calculated using
the conversions EW-\ebv\ from JD.  Although often the EW-\ebv\
calibration of Herbig (1995) is used, JD provide the mean
normalization over four different lines of sight sampling just the diffuse
ISM, instead of only one object (HD 183143), which reduces the effect
of possible peculiarities in using one particular line of sight.
While EW/\ebv\ for individual DIBs can vary factors of several from
sightline to sightline in the data of JD, these variations smooth out
to a scatter of 10-20\% around the mean when a larger sample of DIBs,
such as ours, is considered.  This suggests that calculation of \ebv\
from a sample of DIBs gives a good indication of the minimum
contribution to the total reddening along the sightline by the diffuse
medium (cf. Herbig 1995).

We note in passing that Herbig's EW values for the DIBs in HD 183143 are
larger than those listed by JD.  In fact, JD also note that the EW
values of HD 183143 presented by Herbig (1975) and Herbig \& Leka (1991)
`may' be systematically larger by 20\%.  Le Bertre \& Lequeux (1993)
also mention that their measurements of the DIBs of HD 183143 `possibly
give smaller equivalent widths' than Herbig.  We find the same in our
own high resolution UES spectrum of HD 183143 (courtesy of Ton
Schoenmaker and Eric Bakker).

\begin{table*}
\centering
\caption{\,\, 
The DIBs in the spectra of the 5 massive YSOs and the control star HD
154445. The first two columns identify the DIBs by their wavelength
and their respective normalization to unit \ebv\ as published by
JD. Errors on the equivalent width are given between brackets.  The
value for the $\lambda$6203 DIB is the sum of the 6203 and 6204 \ang\
components resolved by JD, but not by us. NO: Not observed, B: blend
with emission line, an idication of the EW is given NP: Not present --
in general an upper limit of 10m\ang\ is found.
\label{tab:dibs}}
\begin{tabular}{lrrrrrrrrrrrrrrrrrrr}
\hline
\hline

$\lambda$ & W/\ebv  & HD 200775 & HD 154445 & \bd & MWC 1080 & MWC 297 & MWC 349A \\
\ang      &  m\ang \\
\hline
4428 &  2231 &  330 (   70)&  890 (   80)& 1070 (  270)& 1740 (  450)& 1790 (  360)&   3420 (  480) \\
4726 &   123 &   10 (    5)&   65 (   10)&   95 (   10)&  155 (   15)&  315 (   55)&    500 (  100) \\
4963 &    16 &   NP       &    7 (    4)&   20 (    5)&   26 (    2)&   75 (   20)&     NO        \\
5849 &    48 &   NP       &   30 (   10)&   55 (   12)&   75 (   10)&  115 (   20)&    130 (   25) \\
6089 &    17 &   NP       &   15 (    5)&   15 (    5)&   20 (    3)&   25 (    5)&     50 (   10) \\
6195 &    61 &   20 (   10)&   35 (    5)&   50 (   10)&   75 (   10)&   25 (    5)&    130 (   30) \\
6203 &   296 &   25 (   10)&   65 (   10)&  180 (   25)&  200 (   40)&   60 (   25)&    330 (   40) \\
6269 &   137 &   NP       &   50 (   10)&   85 (   10)&   90 (   15)&   35 (   10)&     40 (   25) \\
6376 &    26 &   NP       &   16 (    4)&   20 (    5)&   32 (    6)&   10 (    5)&     95 (    8) \\
6379 &    78 &   12 (    5)&   57 (    7)&   60 (    6)&   93 (   10)&   70 (   10)&    170 (   15) \\
6613 &   231 &   18 (    5)&  120 (   15)&  150 (   20)&  275 (   35)&  115 (   15)&    425 (   45) \\
6660 &    51 &   NP       &   14 (    7)&   17 (    4)&   50 (   20)&   25 (    5)&     65 (    5) \\
6843 &    27 &    6 (    3)&  NP        &   22 (   10)&   25 (    5)&    8 (    2)&     38 (    5) \\
6993 &   116 &   50 (   10)&   75 (    6)&   80 (   10)&   65 (   25)&   35 (   10)&    280 (   35) \\
7119 &    40 &    8 (    5)&   22 (    5)&   30 (    4)&   15 (    6)&   13 (    3)&     58 (    6) \\
7224 &   259 &   NP        &   60 (    7)&  160 (   20)&  140:  (B)  &   35 (   10)&    240: (B) \\
7357 &    48 &   NP        & NP          &   10 (    5)&   30 (   10)& NP          &     20 (    5) \\
7367 &    42 &   NP        & NP          &   20 (    5)&   45 (    5)&   20 (    5)&    110 (   15) \\

\hline
\hline
\end{tabular}
\, \\

\end{table*}

\begin{figure*}
\mbox{\epsfxsize=0.9\textwidth\epsfbox[50 175 470 440]
{
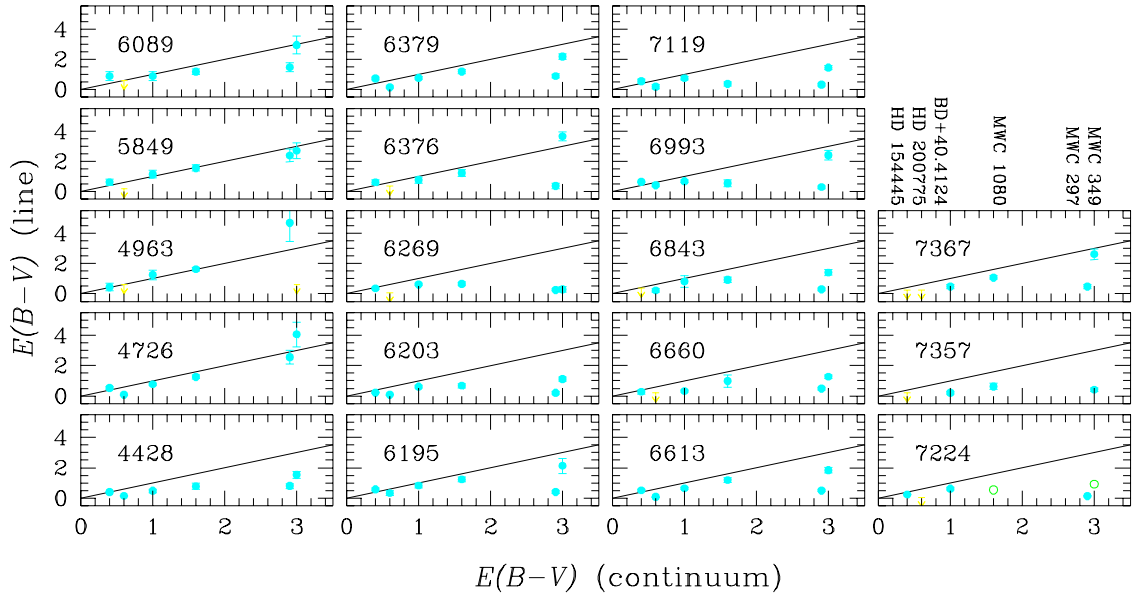}}
\caption{
\,\, Deduced \ebv\ from the measured EWs of the lines for the
different objects plotted per line. The open circles represent 
the bands affected by emission lines.
The solid lines indicate the \ebv\
(DIB) equals the \ebv\ (continuum) relation.
\label{comp1a}
}
\end{figure*}

\subsection{Results}

The results are summarized in Fig.~\ref{comp1a}, where the inferred
\ebv\ of the DIBs
is plotted against the continuum \ebv\ listed in Table~\ref{tab:ebv}.
Note the panel showing the relation for the $\lambda$5849 line.  It is
clear that this line results in deduced \ebv\ values for every star
(except for HD 200775 where the line is not detected) which are in
agreement with the continuum-determined \ebv. The spectra around this
line are shown in Fig.~\ref{fig:dibs}.  Our data strengthen the
tentative suggestion of Chlewicki \ea\ (1986) that the $\lambda$5849
line is frequently a good indicator for the {\it total} extinction
towards an object.  A crude match between line-\ebv\ and
continuum-\ebv\ also seems to exist for the \lam\lam4963, 4726 bands.
However, in most other lines, the trend is that the line-\ebv\ value
is lower than that determined from the continuum.

\begin{figure}
\mbox{\epsfxsize=0.45\textwidth\epsfbox[58 167 544 674]
{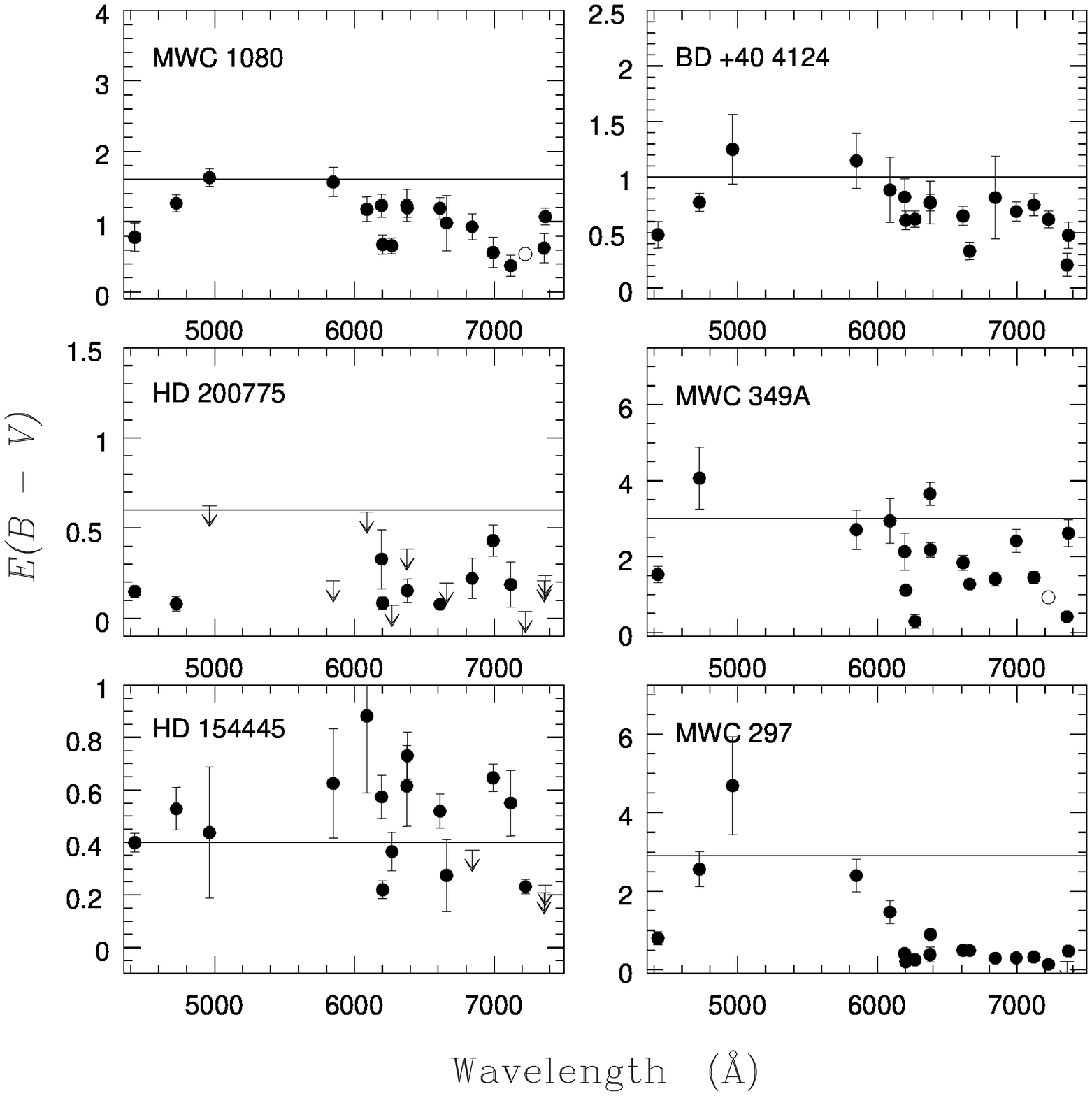}}
\caption{
Inferred \ebv\  from the DIBs on a line by line basis. The horizontal lines represent the
continuum \ebv.
\label{ewrr}
}
\end{figure}

An illustrative plot is provided in Fig.~\ref{ewrr}. It shows the
inferred \ebv\ of the measured lines against DIB wavelength.  The
horizontally drawn lines indicate the continuum \ebv\
(Table~\ref{tab:ebv}).  In especially \bd\ and the standard star HD
154445, the DIBs return an \ebv\ close to the value of the total \ebv,
but in the other cases it can be grasped straightaway that the DIB
\ebv\ is significantly smaller than that determined for the continuum.
The three DIBs in the spectrum of MWC 297 that are sufficiently strong
to trace the total \ebv\ are the same three mentioned above as
correlating well with the total \ebv.  All the other DIBs return
values very much smaller than the continuum \ebv.

The mean  DIB-\ebv, calculated from all detected lines is presented
in Table~\ref{simple}.  In order to avoid a bias towards the larger
EWs, the mean has been calculated using a 1/$\sigma$ weighting,
instead of weighting the measurements according to the inverse square
of their errors (the officially commended weighting scheme). We did
not take into account any non-detections, but it can be seen in
Fig.~\ref{ewrr} that the resulting upper limits on inferred \ebv\
values are mostly in agreement with the means obtained.

In the only instance where we are certain that the diffuse ISM is the 
principal extinction medium, HD 154445, the DIB \ebv\ is in pleasing 
agreement with the continuum \ebv.  For the massive YSOs, it appears that 
the DIB-\ebv\ is always smaller than the continuum \ebv, but that the extent
of the shortfall differs substantially from object to object: specifically,
for \bd\ the DIB-\ebv\ is about 65\% of the continuum \ebv, while the
other extremes are MWC 297 and HD 200775, where the DIBs seem to trace
only 20\% or less of the total \ebv.

\begin{table}
\caption{\ebv\ as derived from the DIBs. n is the number of lines used
in the calculation
 \label{simple} }
\begin{tabular}{llllllll}
\hline
\hline
            & Cont. & & DIB   \\
Star       & \ebv   & n   & $<$\ebv$>$ \\
\hline
HD   154445 & 0.4 &15 &  0.43  ( 0.07) \\
HD  200775  & 0.6 & 9 &  0.14  ( 0.07) \\
\bd         & 1.0 &18 &  0.64  ( 0.08) \\
MWC 1080    &1.6  &17 &  1.01  ( 0.10) \\
MWC 297     &2.9  &18 &  0.47  ( 0.08) \\
MWC 349A    &3.0  &16 &  1.56  ( 0.11) \\

\hline
\hline
\end{tabular}
\end{table}

\section{The contribution of foreground reddening to the total excess}
\label{foreg}

Now that the total \ebv\ and the \ebv\ traced by the DIBs are derived,
it is of interest to have an independent estimate of the extinction due only to
the foreground diffuse interstellar medium along our sightlines.  This is done
by investigating the reddening of  nearby field stars.

\begin{figure}
\mbox{\epsfxsize=0.47\textwidth\epsfbox[25 150 560 690]{
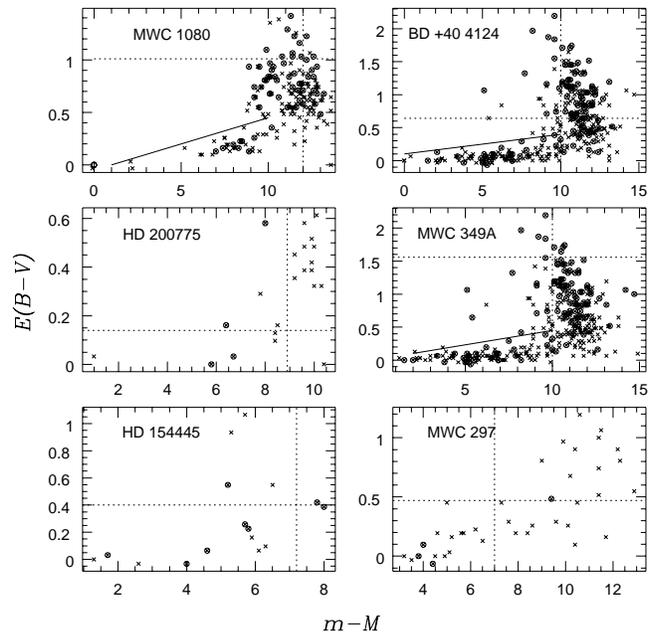}}
\caption{
The relationship between \ebv\ and distance modulus of objects around
the position of the target stars.  The circles indicate objects within
200 arcmin, the crosses represent objects within 400 arcmin.  The
locations of the target stars are indicated by the intersection of
their DIB-\ebv\ and distance. Note that the total \ebv\ to the objects
is in all cases larger than the DIB-\ebv\ (Table~\ref{simple}).The solid
lines represent a conservative estimate of the contribution of the
foreground reddening as a function of distance.
\label{fore}
}
\end{figure}

To facilitate this, we have searched the polarization catalogue compiled by
Matthewson \ea\ (1978), which contains, in combination with
polarization data, values for the extinction and photometric distances
of more than 7500 objects.  For each of our stars, we selected all
objects within a radius of 400 arcmin. The results of this exercise
are set out in Fig.~\ref{fore}, where the \ebv\ for each star is plotted against
its distance modulus {\it m--M}. A consistency check was made by a
comparison with similar relations presented by FitzGerald (1968) and
Neckel \& Klare (1980), and by calculating the \ebv\ and distance
modulus using the spectral types, intrinsic {\it B--V}, and {\it
M$_{V}$} (using \Rv = 3.1) listed by Schmidt-Kaler (1982) for randomly
selected objects.  This showed that the listed \ebv\ and {\it m--M} in
the Matthewson \ea\ (1978) catalogue are in general correct to within
0.2 and 0.5 magnitudes respectively.

In the figure, the objects are indicated with vertical  lines at their estimated distances
\footnote{The adopted distances to the objects are as follows:
MWC 297 at 250 pc based on its spectral type (Drew \ea\ 1997);
MWC 349A - at least 1000 pc, based on spectral type of companion MWC
349B (Cohen \ea\ 1985) ;
HD 200775, 600 pc by Rogers, Heyer \& Dewdney (1995) based on spectral
type B2.5IV. 
\bd, 1000 pc, Hillenbrand \ea\ 1995;
MWC 1080, 2500 pc, Cant\'o \ea\ (1984), but is likely to be smaller - see 
Hillenbrand \ea\ (1982), who cite 1000 pc.  
HD 154445, 275 pc, based on spectral type.
}   and horizontal lines representing  the DIB-\ebv.
The lines of sight to \bd, MWC 349A and MWC 1080 initially show a
trend between the distance modulus and \ebv, which then breaks down at
large distances, where the objects are scattered in \ebv .  Since all
these objects are located in, or near, molecular clouds the scatter is
real, and reflects the many stars at different depths and locations
within patchy clouds.  Where feasible, a conservative upper bound to
the foreground reddening in the lines of sight is indicated by the
solid lines.

 It is not ruled out that the sightlines to MWC 297 and HD 200775 show
 a similar behaviour to that seen in e.g. MWC 349A, but the sparse
 number of data points makes it hard to recognize this.  For example,
 MWC 297 is known to be located within the Aquila Rift at a distance
 of about 250 pc (Dame \& Thaddeus, 1985, Drew \ea\ 1997). However,
 one may see that it is only beyond $\sim$500 parsec ({\it m--M}
 $\sim$ 8.5) that the data points scatter in a similar way as for MWC
 349A.  These stars must be behind the Aquila Rift. It would appear
 that the catalogue of Matthewson \ea\ includes too few stars that are
 located in the Rift at {\it m--M} $\sim$ 6-7.  Similarly, it seems in
 the figure for HD 200775 (positioned within NGC 7023) that \ebv\ only
 starts to become significant beyond {\it m--M} $\sim$ 6.

It is clear that in all cases the foreground extinction is much less
than the total extinction towards the objects (Fig.~\ref{fore},
Table~\ref{simple}). For the \ebv\ traced by the DIBs some interesting
differences appear.  For both MWC 297 and HD 200775 the foreground
extinction may be comparable with the \ebv\ as calculated from the
DIBS, but not larger.  In the other cases, the DIB-\ebv\ is
intermediate between the total \ebv\ and the foreground extinction.
This suggests that the molecular clouds, or even the circumstellar
material, harbour conditions in which the DIB carriers can be present
albeit at reduced abundance.

We remind however that there is a small collection of DIBs in the
spectrum of MWC 297 that is particularly strong; namely $\lambda$5849,
$\lambda$4963 and $\lambda$4726 appear to trace the total continuum
\ebv.

\section{Concluding remarks}

DIB features are in general well-correlated with the interstellar
extinction for field objects, but deviate in instances where
circumstellar or parental cloud material is present.  The results in
Table~\ref{simple} and in the previous section certainly indicate that
the latter is true for our sample of massive YSOs as well.
The diffuse ISM contributes only a fraction of the total reddening
towards the  YSOs -- the rest of the extinction being
due to either cloud or circumstellar material.
 
In conclusion, our study of Diffuse Interstellar Bands towards a
sample of massive YSOs can be summarized as follows:

\begin{enumerate}

\item{ 
From our crude
estimates of the foreground extinction we find that a large fraction
of the extinction to these stars is not provided by the diffuse
interstellar medium, but by circumstellar material or dust within the
parental cloud. 
}

\item{ 
By and large, the DIB-\ebv\ is smaller than the total \ebv\ towards
the optically visible massive YSOs - but is mostly larger than the
expected foreground reddening. This means that at least a fraction of
the DIBs are formed within the molecular clouds, but less efficiently
than in the normal diffuse ISM.  This implies that conditions in this
material are such that DIB excitation is weaker or that the DIB
carriers are less efficiently formed or more prone to destruction.
The DIBs trace most of the \ebv\ suffered by \bd\ but very
little in the line of sight to MWC 297.  Such variations in
`efficiency' clearly imply that the conditions vary strongly from one
cloud of massive star formation to the other.
}

\item{
If the DIBs are due to different types of carrier, the strength of the
lines will be proportional to the respective column densities of the
carriers. The effective normalization of line-strengths to \ebv\ for
`normal' diffuse interstellar material then gives us a handle on their
relative abundances.  From the fact that the \lam5849 line seems to
trace the total \ebv, one may infer that the carrier of this line is
more readily formed or less likely to be destroyed,  in star-forming  clouds 
than other lines.   The same may also be true for the $\lambda$4726 and  
$\lambda$4963 DIBs as strikingly illustrated by the case of MWC 297.
}

\end{enumerate}

\paragraph*{\it Acknowledgments}
Kaylene Murdoch and Phil Lucas are thanked for obtaining the observations.
The allocation of time on the WHT was awarded by PATT, the United
Kingdom allocation panel.  This work was entirely funded by the
Particle Physics and Astronomy Research Council of the United Kingdom.

 
\newcommand{\apspsc}[1]{{ Astrophys. Sp. Sci.}, { #1}} 
\newcommand{\aj}[1]{{ AJ}  { #1}} 
\newcommand{\apj}[1]{{ ApJ}  { #1}} 
\newcommand{\apjlett}[1]{{ ApJL}  { #1}} 
\newcommand{\apjsupp}[1]{{ ApJS}  { #1}} 
\newcommand{\aanda}[1]{{ A\&A}  { #1}} 
\newcommand{\aandasupp}[1]{{ A\&AS}  { #1}} 
\newcommand{\araa}[1]{{ ARA\&A}  { #1}} 
\newcommand{\pasp}[1]{{ PASP}  { #1}} 
\newcommand{\nature}[1]{{ Nature}  { #1}} 
\newcommand{\mnras}[1]{{ MNRAS}  { #1}} 
\newcommand{\sva}[1]{{ SvA}  { #1}} 
\newcommand{\irphys}[1]{{ Infrared Phys. Technol.}, { #1}}

\end{document}